\documentclass[10pt, conference]{IEEEtran}
\IEEEoverridecommandlockouts

\usepackage[a4paper,
            left=0.670in,
            right=0.670in,
            top=0.750in,
            bottom=1.05in]{geometry}

\usepackage{algorithmic}
\usepackage{url}
\def\BibTeX{{\rm B\kern-.05em{\sc i\kern-.025em b}\kern-.08em
    T\kern-.1667em\lower.7ex\hbox{E}\kern-.125emX}}
\usepackage[hidelinks]{hyperref}
\hypersetup{colorlinks=false}
\usepackage{cite}
\usepackage{amsmath,amssymb,amsfonts}
\usepackage{graphicx,color}
\usepackage{textcomp}
\usepackage{xcolor}

\usepackage[ruled,vlined,linesnumbered]{algorithm2e}
\DontPrintSemicolon
\SetKwInput{Input}{Input}
\SetKwInput{Output}{Output}
\SetKw{To}{ to }
\SetKw{Print}{Return }

\usepackage{orcidlink}
\usepackage{soul}
\usepackage[caption=false,font=footnotesize]{subfig}
\usepackage{float}
\usepackage{placeins}
\usepackage{adjustbox}
\usepackage{multirow}

\begin{document}

\title{SmartUT: Receive Beamforming for Spectral Coexistence of NGSO Satellite Systems\\ 
\thanks{This research was funded by the Luxembourg National Research Fund (FNR) under the project SmartSpace (C21/IS/16193290).}}



\author{\IEEEauthorblockN{Almoatssimbillah Saifaldawla, Eva Lagunas, Flor Ortiz, Abuzar B. M. Adam and Symeon Chatzinotas}
\IEEEauthorblockA{\textit {Interdisciplinary Centre for Security Reliability and Trust (SnT),}\\
\textit{University of Luxembourg, Luxembourg} \\e-mail: \{moatssim.saifaldawla, eva.lagunas, flor.ortiz, abuzar.babikir, symeon.chatzinotas\}@uni.lu
} }

\maketitle

\begin{abstract}

In this paper, we investigate downlink co-frequency interference (CFI) mitigation in non-geostationary satellite orbits (NGSOs) co-existing systems. Traditional mitigation techniques, such as Zero-forcing (ZF), produce a null towards the direction of arrivals (DOAs) of the interfering signals, but they suffer from high computational complexity due to matrix inversions and required knowledge of the channel state information (CSI). Furthermore, adaptive beamformers, such as sample matrix inversion (SMI)-based minimum variance, provide poor performance when the available snapshots are limited. We propose a Mamba-based beamformer (MambaBF) that leverages an self-supervised deep learning (DL) approach and can be deployed on the user terminal (UT) antenna array, for assisting downlink beamforming and CFI mitigation using only a limited number of available array snapshots as input, and without CSI knowledge. Simulation results demonstrate that MambaBF consistently outperforms conventional beamforming techniques in mitigating interference and maximizing the signal-to-interference-plus-noise ratio (SINR), particularly under challenging conditions characterized by low SINR, limited snapshots, and imperfect CSI.
\end{abstract}


\section{INTRODUCTION}
Satellite communications (SatCom) will play a vital role in next-generation wireless networks by providing service to vast areas that lack terrestrial network coverage, especially with the rapidly growing Low-Earth orbit (LEO) mega-constellations \cite{Hraishawi23}. These constellations are distributed in the Non-Geostationary Satellite orbits (NGSOs) and must manage co-existence strategies for sharing the limited spectrum \cite{He2023}. Despite the proactive coordination efforts established by the International Telecommunication Union (ITU) among NGSO operators \cite{itu22}, there remains a risk of unintentional co-frequency interference (CFI), especially in the Ku/Ka bands where tens of thousands of NGSOs operate, potentially causing severe quality of service (QoS) degradation \cite{Braun19}. Thus, advanced interference avoidance and mitigation techniques that ensure both low complexity and efficient mitigation of unintentional interference in NGSO spectrum co-existence scenarios are needed. 

Planar arrays are becoming popular user terminals (UT) equipments, because they can be used to steer the receiver electronically towards the moving satellite line of sight \cite{Starlink96:online}. This work focuses on the adaptive beamforming (ABF) techniques on the UT side, where low complexity and fast adaptation are needed (due to the time-varying nature of the scenario). The goal is to generate antenna weights to accurately steer the main lobe towards the signal of interest (SOI) locations while minimizing the antenna gain towards interference locations. 

Conventional correlation-based ABF methods, such as the Capon beamformer, also known as Minimum Variance Distortionless Response (MVDR) \cite{van2002}, rely on knowledge of the covariance (or correlation) matrix of the received signals. The covariance matrix is typically estimated using snapshots of the received signals, and accurate computation of the inverse covariance matrix is key to nulling interferers while preserving the SOI. Furthermore, this approach relies on channel state information (CSI) to constrain the beamformer towards the SOI direction. Moreover, traditional ABF's main drawbacks are the high computational complexity involved in the weight calculation and the iterative nature of the algorithms, which makes the temporal response of the fast-moving satellites a critical concern. 

Deep learning (DL)-based beamforming techniques have emerged as a potential solution to the traditional beamforming techniques' drawbacks \cite{Kass2022}. DL-based beamforming typically replaces or augments the classical weight calculation with a neural network that directly outputs beamforming weights. In this context, the authors in \cite{sho2024} compared the performance of an encoder-based beamformer, a convolutional neural network (CNN) and MVDR approaches in terms of signal-to-interference-plus-noise ratio (SINR) for LEO satellites communication systems; these approaches rely on supervised learning, which requires per definition input-output pairs. Obtaining such labeled data can be challenging due to the unpredictable environments of NGSOs, making the ground truth difficult to ascertain. In \cite{Cemil2024}, the authors introduced an unsupervised deep neural network (DNN) beamforming (NNBF) model for the terrestrial uplink sum-rate maximization problem; the CSI was used as an input, but in the CFI case, only the desired CSI can be estimated. Furthermore, such a model might be sensitive to imperfect CSI scenarios. 

\begin{figure}[ht]
    \centering{
    \includegraphics[width=3.0in]{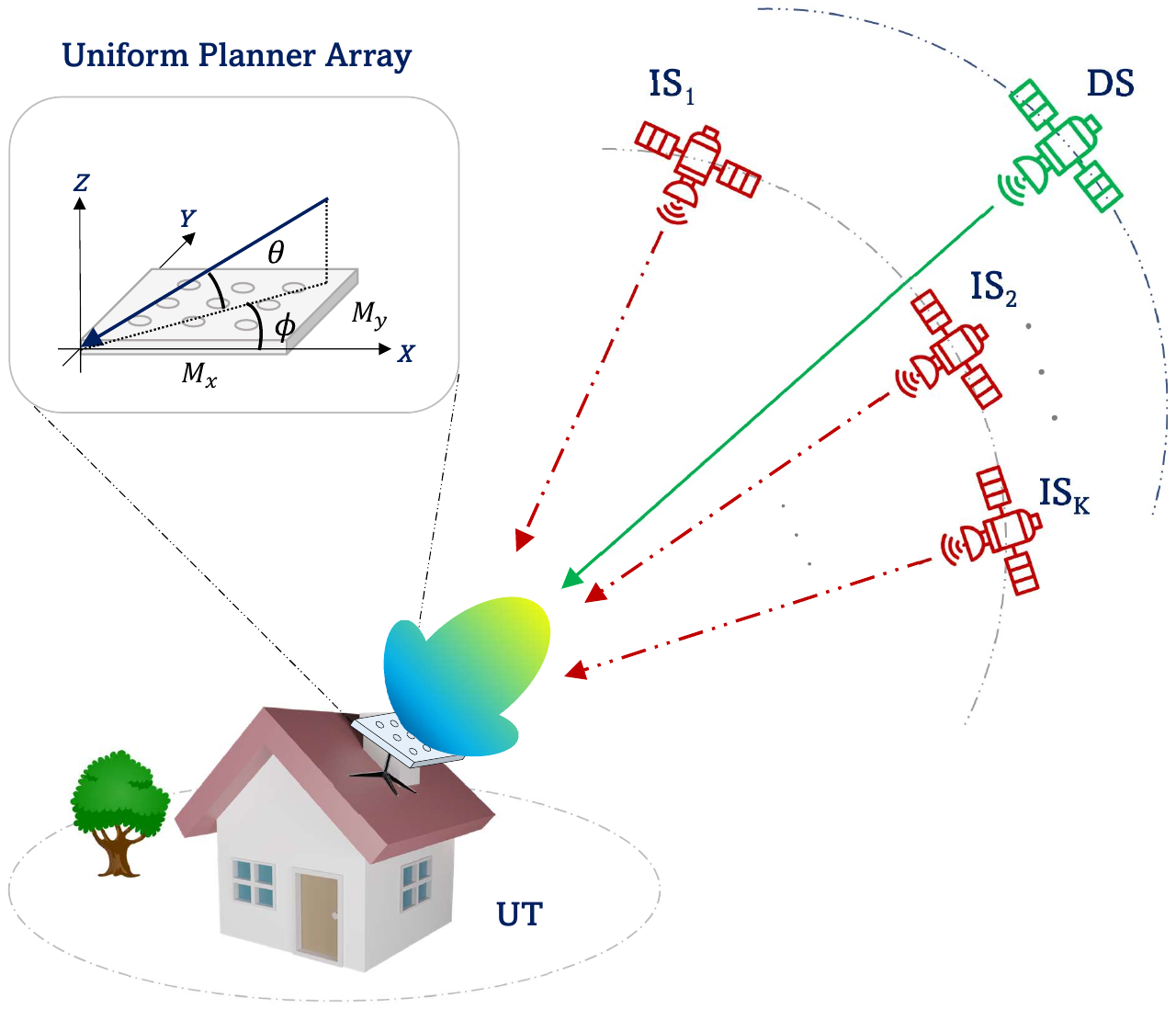}}
    \caption{Desired satellite downlink (green) with aggregated interference from a second independent NGSO constellation (red).}
    \label{fig:sysmod}
\end{figure}

In this paper, we propose an self-supervised DL-based beamformer approach using Mamba state space model (SSM) layers \cite{gu2024mamba}. These layers effectively capture both spatial and temporal features from the input data that may be relevant to generating the correct beamforming weight vectors. Mamba-based beamformer (MambaBF) can be deployed on the UT side for more energy-efficient training and inference rather than on-board training \cite{10615457}, for assisting downlink CFI mitigation in NGSO communications. MambaBF is trained using the available received array snapshots as input to produce a beamforming weight vector that can maximize SINR by nulling interference directions.

\section{SYSTEM MODEL}

We investigate CFI between two coexisting NGSO satellite constellations, as illustrated in Fig.\ref{fig:sysmod}. In particular, we consider a desired satellite ($\mathrm{DS}$) serving a fixed UT, and some interfering satellites ($\mathrm{ISs}$) serving users in close vicinity of the UT location and therefore, pointing their signal power towards the nearby UT location. Our focus in this work is on the UT-side receive beamforming strategy, designed to avoid unintentional interference from other satellites.

\subsection{Channel and signal models}

Consider the uniform planar array (UPA) at the UT lying in the $xy$-plane with $M_x \times M_y$ antenna elements on the $x$ and $y$ axes respectively, and the array size $M = M_x M_y$. Therefore, the UT array response vector can be written as \cite{Ayach14},

\begin{equation}
\begin{aligned}
\mathbf{v}(\phi, \theta) &=  \frac{1}{\sqrt{M}} \left[ 1, \cdots, e^{j \frac{2 \pi}{\lambda} a (m_x \sin \phi \cos \theta + m_y \sin \theta)}, \cdots, \right. \\
&\left. e^{j \frac{2 \pi}{\lambda} a ((M_x-1) \sin \phi \cos \theta + (M_y-1) \sin \theta)} \right]^{T} \in 
 \mathbb{C}^{M},
\end{aligned}
\label{eq:UTresponse }
\end{equation}
where $a$ is the inter-element spacing, $0 \leq m_x < M_x$ and $0 \leq m_y < M_y$ are the antenna indices in the $xy$-plane, $\phi$ and $\theta$ are the azimuth and the elevation angles respectively, representing the directions of arrival (DoAs). For notational brevity, we define a DoA as $\varphi \triangleq (\phi, \theta )$.

We are modelling a scenario where the $\mathrm{DS}$ transmits a signal ($s_x \in 
 \mathbb{C} $) with $\mathbb{E}\{s_x s_x^H\} = 1$ to its UT. Eventually, $K (K < M)$ interfering satellites ($\mathrm{IS}_{k}, \quad  k \in 1, \cdots, K$) are visible to the UT and unintentionally transmit an interference signal ($s_{i, k} \in 
 \mathbb{C}$) towards it (the underscore "$i$" and "$d$" will be used to indicate interference satellite and desired satellite parameters respectively). When placing the UT in an obstacle-free elevated space as recommended by many operators (e.g., STARLINK \cite{Starlink96:online}), the line-of-sight (LoS) component becomes dominant, thus the non-LoS (NLoS) propagation components can be ignored. Assuming $\mathrm{DS}$'s Doppler shift is compensated for at the UT, we define the statistical time-varying channel state information (sCSI) model between the $\mathrm{DS}$ satellite and the UT at time instant $t$ and frequency $f_{c, d}$ as follows\cite{You2020}, 

\begin{equation}
    \mathbf{h}_d =\chi_d \cdot \mathbf{v}_{d} (\varphi_d) \in 
 \mathbb{C}^{M}, \label{eq:dChannel}
\end{equation}
where $\chi_d = \sqrt{P_d L_d G_d^{ut}(\varphi_d)}$ is the channel gain, and $P_d$, $G_d^{ut}(\varphi_d)$  are the $\mathrm{DS}$ Equivalent Isotropic Radiated Power (EIRP), and the maximum UT receive gain at $(\varphi_d)$ direction. $L_d = (\frac{\lambda_d}{4\pi r_d})^2$ is the free space path loss inverse, $r_d$ is the slant range between UT and $\mathrm{DS}$, and $\lambda_d$ is the wavelength. Accordingly, the effective channel vector for the $k^{th}$ interfering satellite $\mathrm{IS}_{k}$ to UT receiver can be derived as,

\begin{equation}
    \mathbf{h}_{i, k} = \chi_k \cdot \mathbf{v}_{k} (\varphi_{k}). \label{eq:iChannel}
\end{equation}
where $\chi_k = \sqrt{P_k L_k G_k^{ut}(\varphi_k)}$. The UT array receiving gain $G^{ut}(\varphi)$ can be obtained as follows, 

\begin{equation}
    G^{ut}(\varphi) = \eta^{ut} D^{ut}(\varphi). \label{eq:ArrayGain}
\end{equation}
where $\eta^{ut}$, $D^{ut}$ are the UPA antenna efficiency and the directivity towards $\varphi$ direction. Note that the channel model adopted in (\ref{eq:dChannel}) and (\ref{eq:iChannel}) is applicable over a specific coherence time "$t$" where the relative positions of the satellites and UT do not change significantly, and the physical channel parameters remain invariant. In this work, we simulate a constitutive sequence of coherence times for the sake of adaptability. Assuming that the UT array at the time instant $t$ can capture $L$ total snapshots, the $l^{th}$ snapshot vector can be expressed as,
\begin{equation}
    \mathbf{y}[l] = \mathbf{h}_d s_x[l]  +  \mathbf{H}_i \mathbf{S}_{i}[l]  + \mathbf{n}[l] \quad \in 
 \mathbb{C}^{M}, \label{eq:UTsignal}
\end{equation}
where $\mathbf{H}_i \triangleq \left[ h_{i, 1}; h_{i, 2}; \dots;h_{i, K}   \right] \in 
 \mathbb{C}^{M \times K}$, and  $\mathbf{S}_{i} \triangleq \left[ s_{i, 1}; s_{i, 2}; \dots;s_{i, K}   \right] \in 
 \mathbb{C}^{K}$, here $\mathbf{n} \in \mathbb{C}^{M} $ presents the additive white Gaussian noise (AWGN) at the UT receiver with i.i.d. $\mathcal{CN}(0, \sigma_{n}^2= \kappa T^{ut} BW_d)$, where  $\kappa$, $T^{ut}$, and $B_d$ are the Boltzmann constant, equivalent noise temperature of the UT, and $\mathrm{DS}$ signal bandwidth, respectively.

\subsection{Traditional beamforming techniques}
For the UT to recover $\mathrm{DS}$ signals, we need to design a beamformer \textit{weight} vector $\mathbf{w} \in \mathbb{C}^{M}$ with the objective of steering the UT main beam towards $\varphi_{0}$ direction and eliminating "null" interference coming from $\varphi_{K}$ directions. The estimated signal after ($\hat{s}\in 
 \mathbb{C}$) after applying $\mathbf{w}$ is,
\begin{equation}
    \hat{s}[l] = \mathbf{w}^H \mathbf{y}[l] = \mathbf{w}^H \mathbf{h}_d s_x[l]  +  \mathbf{w}^H \mathbf{H}_i \mathbf{S}_{i}[l]   + \mathbf{w}^H  \mathbf{n}[l]. \label{eq:UTestimate}
\end{equation}

When we know the covariance matrix of a single snapshot $\mathbf{R}[l] \triangleq  \mathbb{E} \{\mathbf{y}[l]\mathbf{y}^{H}[l] \}$, the beamformer $\mathbf{w}$ can be obtained by maximizing the output SINR \cite{van2002},

\begin{equation}
\mathrm{SINR}_{\text{out}}[l] = \frac{ \mathbf{w}^H \mathbf{R}_{d}[l] \mathbf{w} }{\mathbf{w}^H \mathbf{R}_{i+n}[l] \mathbf{w}} , \label{eq:UTSINR}
\end{equation}
where $\mathbf{R}_{d} = \mathbf{h}_d \mathbf{h}_d^H \in \mathbb{C}^{M \times M}$, $\mathbf{R}_{i+n} = \mathbf{H}_i \mathbf{H}_i^H  + \sigma_{n}^2 \mathbf{I}_M  \in \mathbb{C}^{M \times M} $ are the desired covariance matrix and the interference-plus-noise covariance matrix, respectively.

The problem of maximizing ($\ref{eq:UTSINR}$) is mathematically equivalent to the MVDR beamforming problem \cite{godara2004smart}. In particular, the MVDR beamformer minimizes UT array input power while maintaining unity gain in $\mathrm{DS}$ direction,
\begin{equation}
\mathbf{w}_{\text{MVDR}} = \arg \min_{\mathbf{w}} \mathbf{w}^H \mathbf{R} \mathbf{w} \quad \text{subject to} \quad \mathbf{w}^H \mathbf{v}_d = 1,\label{eq:mvdrbeamformer}
\end{equation}

This optimization problem leads to the following weights for the beamformer $ \mathbf{w}_{\text{MVDR}} = \frac{\mathbf{R}^{-1} \mathbf{v}_d}{\mathbf{v}_d^H \mathbf{R}^{-1} \mathbf{v}_d}$. In the finite sample case, \( \mathbf{R} \) is unavailable; it is usually replaced by the sample covariance matrix \( \hat{\mathbf{R}} \) for \( L \) total snapshots, \cite{Ger2010}
\begin{equation}
\hat{\mathbf{R}} = \mathbb{E}\{\mathbf{Y}\mathbf{Y}^H\}  = \frac{1}{L} \sum_{l=1}^{L} \mathbf{y}[l] \mathbf{y}^H[l],
\label{eq:snapshotsCov}
\end{equation}
where $\mathbf{Y} = \left[ \mathbf{y}[1],\mathbf{y}[2], \cdots \mathbf{y}[L] \right] \in \mathbb{C}^{M \times L} $. When $\hat{\mathbf{R}}$ is used to formulate (\ref{eq:mvdrbeamformer}), the solution is then called sample matrix inversion (SMI)-based minimum variance beamformer and is given by $ \mathbf{w}_{\text{SMI}} = \frac{\hat{\mathbf{R}}^{-1} \mathbf{v}_d}{ \mathbf{v}_d^H \hat{\mathbf{R}}^{-1} \mathbf{v}_d }$. As \( L \) increases, \( \hat{\mathbf{R}} \) converges to the theoretical covariance matrix $ \mathbf{R} = \mathbf{R}_{d} + \mathbf{R}_{i+n}$ and the corresponding $\mathrm{SINR}$ will approximate the optimal value as \( L \rightarrow \infty \) . However, as the available snapshots size \( L \) decreases, the gap between \( \hat{\mathbf{R}} \) and \( \mathbf{R} \) increases, which dramatically affects the performance \cite{van2002}. Here $\mathrm{SINR}$ can be measured based on the $L^{th}$ snapshot as a random variable of a discrete random process [see Eq. 7.90 \cite{van2002}], instead, we assume average $\mathrm{SINR}$ ($\mathrm{ASINR}$) from the available snapshots,

\begin{equation}
\mathrm{ASINR}_{\text{out}} = \frac{1}{L} \sum_{l=1}^L{\frac{\mathbf{w}^H \mathbf{R}_{d}[l] \mathbf{w} }{\mathbf{w}^H \mathbf{R}_{i+n}[l] \mathbf{w}}} , \label{eq:UTASINR}
\end{equation}

In addition to adaptive $\mathbf{w}_{\text{SMI}}$, we will compare our results with maximum ratio combining (MRC) and ZF beamformers as follows 

\begin{equation}
\mathbf{w}_{\text{MRC}} = \frac{\mathbf{h}_d}{\| \mathbf{h}_d\|}.\label{eq:MRCbeamformer}
\end{equation}

\begin{equation}
\mathbf{w}_{\text{ZF}} = \frac{\left( \mathbf{I}_M - \mathbf{H}_i (\mathbf{H}_i^H \mathbf{H}_i)^{-1} \mathbf{H}_i^H \right) \mathbf{h}_d}{\| \left( \mathbf{I}_M - \mathbf{H}_i (\mathbf{H}_i^H \mathbf{H}_i)^{-1} \mathbf{H}_i^H \right) \mathbf{h}_d \|} .\label{eq:ZFbeamformer}
\end{equation}

The benchmark approaches required different information to perform well. SMI requires $\mathbf{\hat{R}}$ and $\mathbf{h}_d$, MRC requires $\mathbf{h}_d$, while ZF requires $\mathbf{H}_i$ and $\mathbf{h}_d$. 


In practice, UT might be able to estimate the CSI of the $\mathrm{DS}$ link using a pilot signal or have prior knowledge of the $\mathrm{DS}$ steering vector $\mathbf{v}_d$, but cannot access any information about $\mathrm{ISs}$ links. The channel estimated usually suffers from an estimation error $\mathbf{e}_d$,

\begin{equation}
    \hat{\mathbf{h}}_d = \mathbf{h}_d + \mathbf{e}_d
\end{equation}
where \(\mathbf{e}_d\) can be modeled as a complex Gaussian random variable $ \mathbf{e}_d \sim \mathcal{CN}(\mathbf{0}, \sigma_e^2 \mathbf{I})$.


\section{PROPOSED APPROACH}

\begin{figure*}[ht]
    \centering
    \includegraphics[width= 5.0 in]{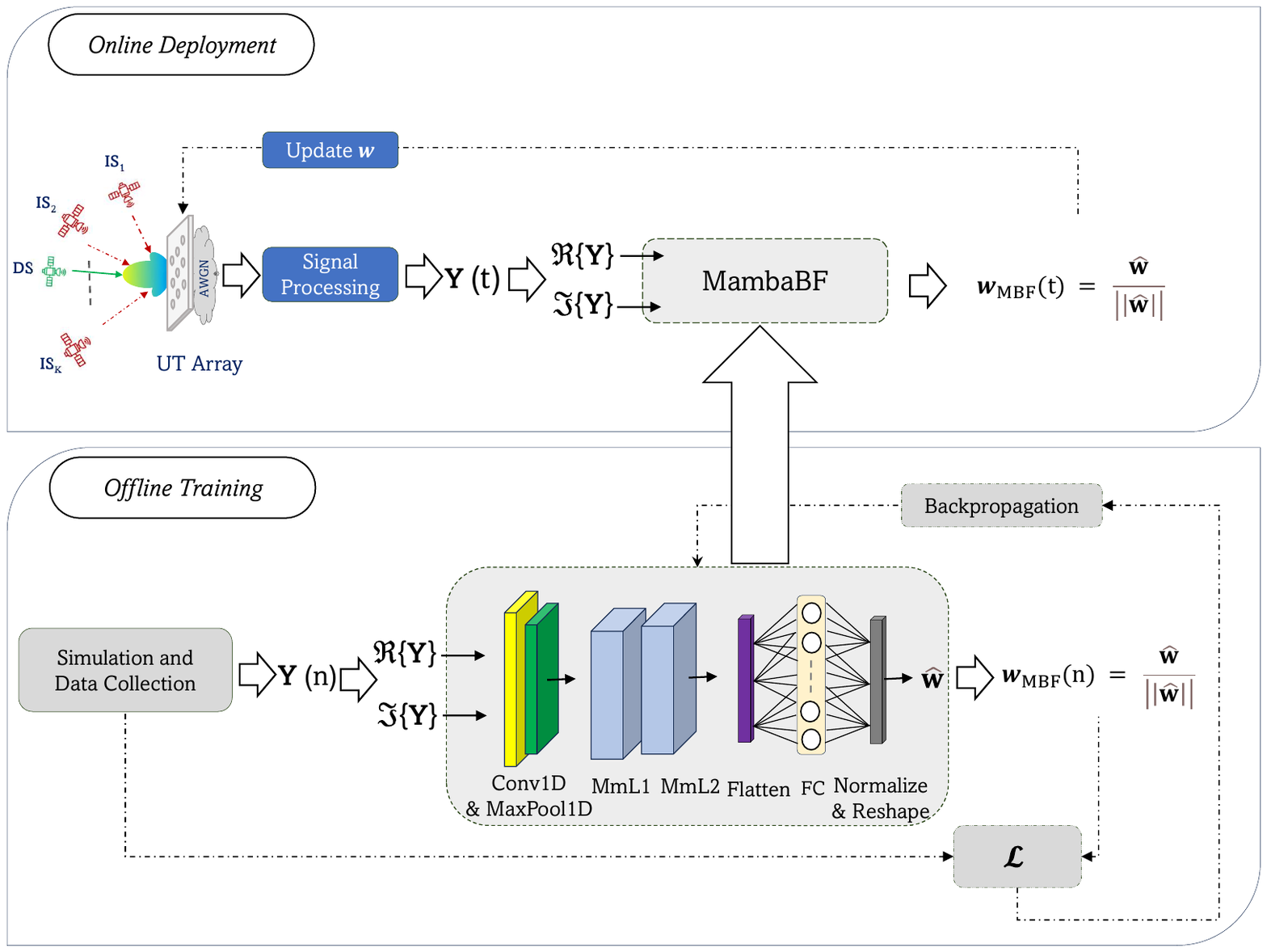}
    \caption{Systematic view of the proposed MambaBF approach, "n" refer to the training sample.}
    \label{fig:dlmod}
\end{figure*}

 To mitigate NGSOs' CFI, we need a mechanism that can be trained offline to utilize any available $L$ snapshots of the array output $\mathbf{Y}$ and generate a robust weight vector $\mathbf{w}_{\text{MBF}}$ without prior knowledge of desired or interference channels. We propose a MambaBF to tackle this problem. Mamba-based SSMs layers were first introduced by \cite{gu2024mamba} with the aim to address Transformers’ computational inefficiency on long sequences of the input data by allowing the parameters of the SSM to selectively retain or discard information along the sequence dimensions based on the current state (i.e., snapshots in our case), thus improving the model's ability to handle sequences effectively. 


\subsection{Model Overview}


We aim at generating $\mathbf{w}_{\text{MBF}}$ directly from the received snapshots $\mathbf{Y}$ signal. Fig.~\ref{fig:dlmod} (lower panel) illustrates the offline training pipeline. When $L$ snapshots are collected, the data can be arranged in a matrix $\mathbf{Y}(n) \in\;\mathbb{C}^{M \times L},$ where $n$ is a training sample. Each complex entry is converted into its real and imaginary parts before being fed into the neural network, giving an input real-valued tensor of shape $(2M, L)$. 

The real and imaginary parts of $\mathbf{Y}$ are first normalized and then processed by a 1D convolution (conv1D) plus max-pooling (MaxPool1D) across the snapshot dimension for dimensionality reduction. Each Conv1D output is \text{Batch}-normalized to stabilize training. After these initial layers, the latent input $\mathbf{x} \in \mathbb{R}^{M_z \times L} $, $M_z < 2M$, where $M_z$ is the latent space size, is passed through the Mamba-based SSM layers (MmLs), followed by fully connected (FC) layers. The final output is a weight vector $\hat{\mathbf{w}} \in \mathbb{R}^{2M}$, reshaped into $\mathbb{C}^M$, then normalized to yield
\begin{equation}
  \mathbf{w}_{\mathrm{MBF}} \;=\;\frac{\hat{\mathbf{w}}}{\|\hat{\mathbf{w}}\|}.
\end{equation}

\begin{figure}[!ht]
    \centering
    \includegraphics[width= 2.5in]{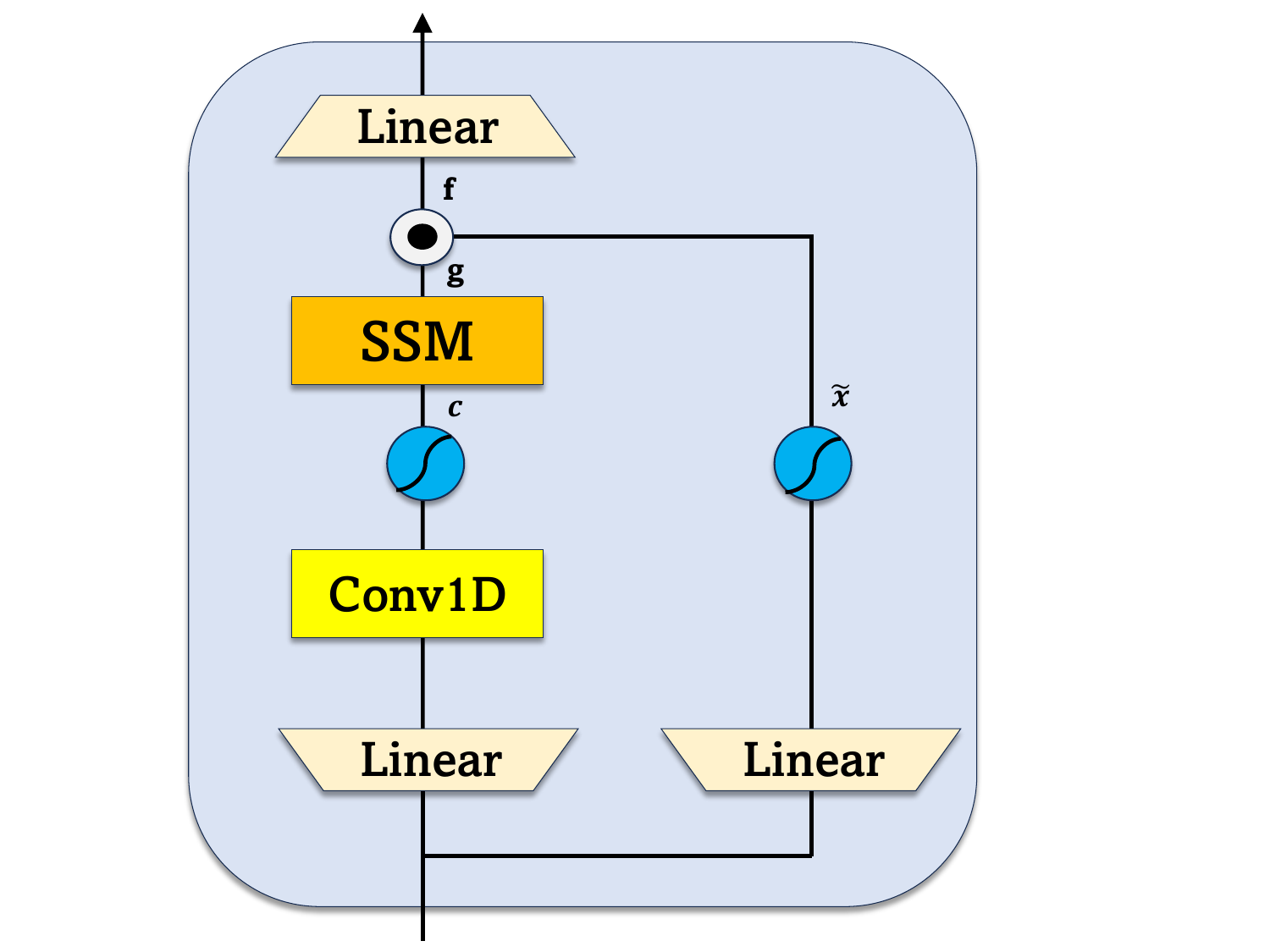}
    \caption{Mamba layer (MmL) }
    \label{fig:dlmodMB}
\end{figure}

\subsection{Core Mamba Layer (MmL).}
The structure of the Mamba layers (MmLs)~\cite{gu2024mamba} incorporates multi-modal feature extraction capabilities through the integration of linear transformations, convolutional operations, and recurrent sequence modeling. We enhance the local and global extraction of temporal and spatial features by stacking two Mamba layers. An MmL block handles sequences by merging two concurrent linear transformations of the input, as illustrated in Fig.~\ref{fig:dlmodMB}. One pathway applies scaled exponential linear unit (SeLU) activation on the input,
\begin{equation}
           \tilde{\mathbf{x}}(\ell)
            \;=\;
            \mathrm{SeLU}\bigl(\mathbf{x}(\ell))\bigr),
\end{equation}
the other linear path is first processed by a conv1D with kernel size $k$, typically performing,
\begin{equation}
            \mathbf{c}(\ell)
            \;=\;
            \mathrm{SeLU}\bigl(\mathrm{Conv1D}_k(\mathbf{x}(\ell))\bigr),
\end{equation}
a Gated Recurrent Unit (GRU) further refines this sequence representation $\mathbf{c}(\ell)$ as a simple SSM model, as it unfolds over $\ell$ modeling long-term dependencies in the data,
\begin{equation}
            \mathbf{g}(\ell)
            \;=\;
            \mathrm{GRU}\bigl(\mathbf{c}(\ell), \mathbf{g}(\ell-1)\bigr),
\end{equation}
the output of the GRU is then element-wise multiplied with the skipped connection SeLU activation output, effectively acting as an attention-like mechanism to emphasize relevant features,
\begin{equation}
            \mathbf{f}(\ell) 
            \;=\; 
            \mathbf{W}_{\mathrm{out}}\!\Bigl(\mathbf{g}(\ell)\,\odot\,\tilde{\mathbf{x}}(\ell)\Bigr),
\end{equation}
where $\mathbf{W}_{\mathrm{out}}$ is the model output projection matrix, $\tilde{\mathbf{x}}(\ell)$ is a skip-connection transform of $\mathbf{x}(\ell)$, and $\odot$ denotes element-wise multiplication, 

\subsection{self-supervised Training Objective}

The proposed learning procedure offers self-supervised training, and the loss function is customized to minimize $\mathcal{L}_{\mathrm{ASINR}}(n) = - \frac{1}{L} \sum_{(n, l)}^{(n, L)} \mathrm{SINR}(l, n)$,  where $n$ is a training sample. Given $\mathbf{w}_{\mathrm{MBF}}(\mathbf{Y}(n))$, we measure $\mathcal{L}(n)$ in terms of the received power from the desired direction vs. interference plus noise. We then perform backpropagation through the MambaBF network using $\mathcal{L}(n+1)$ as our new cost.


\section{SIMULATION RESULTS}

   

   \begin{figure*}
   \centering
   \includegraphics[width=3.0in]{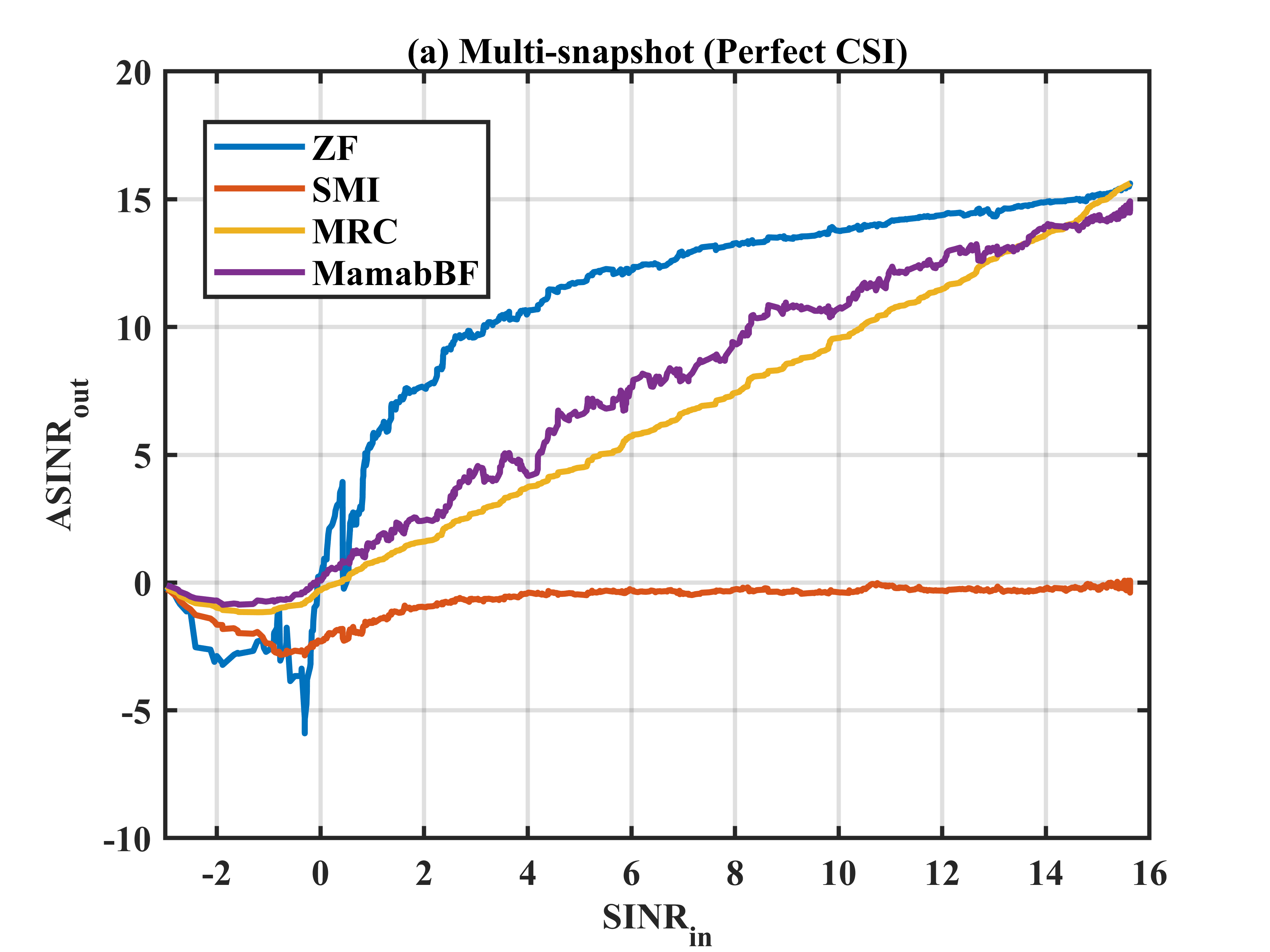}
\includegraphics[width=3.0in]{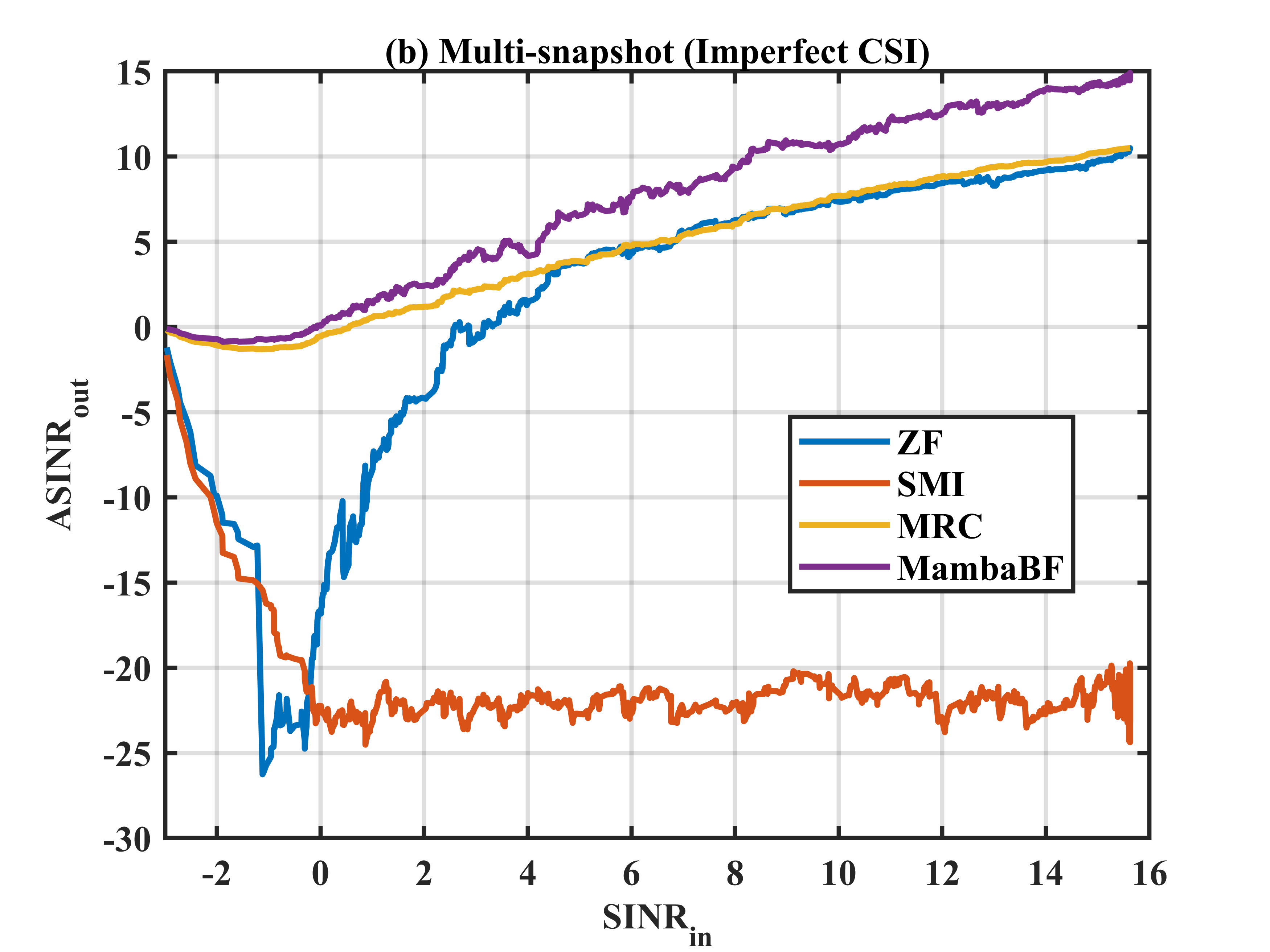}
  \caption{$\mathrm{ASINR}$ for multi-snapshot with (a) perfect CSI (left) and (b) imperfect CSI (right)}
  \label{fig:MS_PImCSI}
\end{figure*}

\begin{figure}[ht]
  \centering
  \includegraphics[width=2.0in]{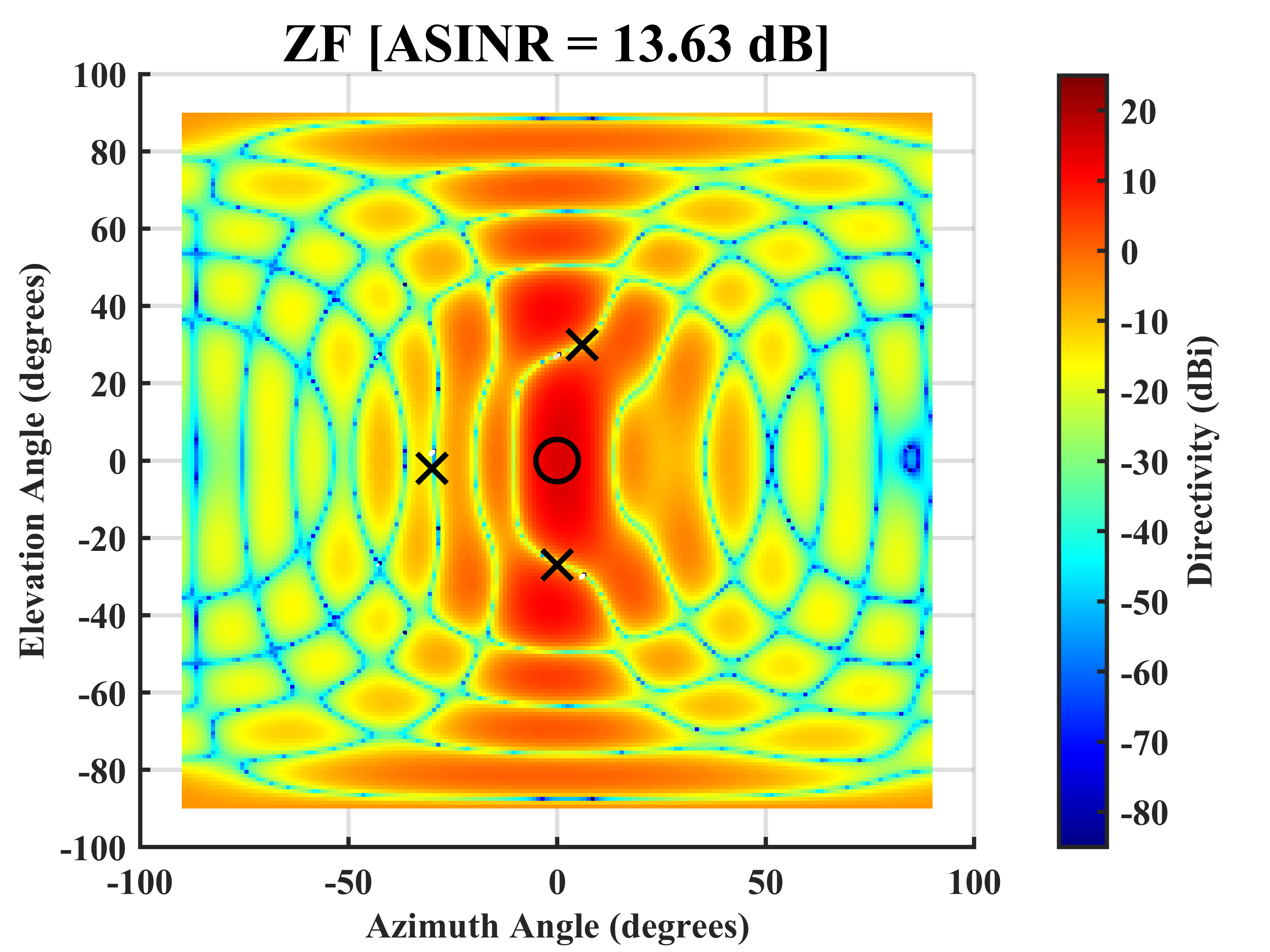}
  \includegraphics[width=2.0in]{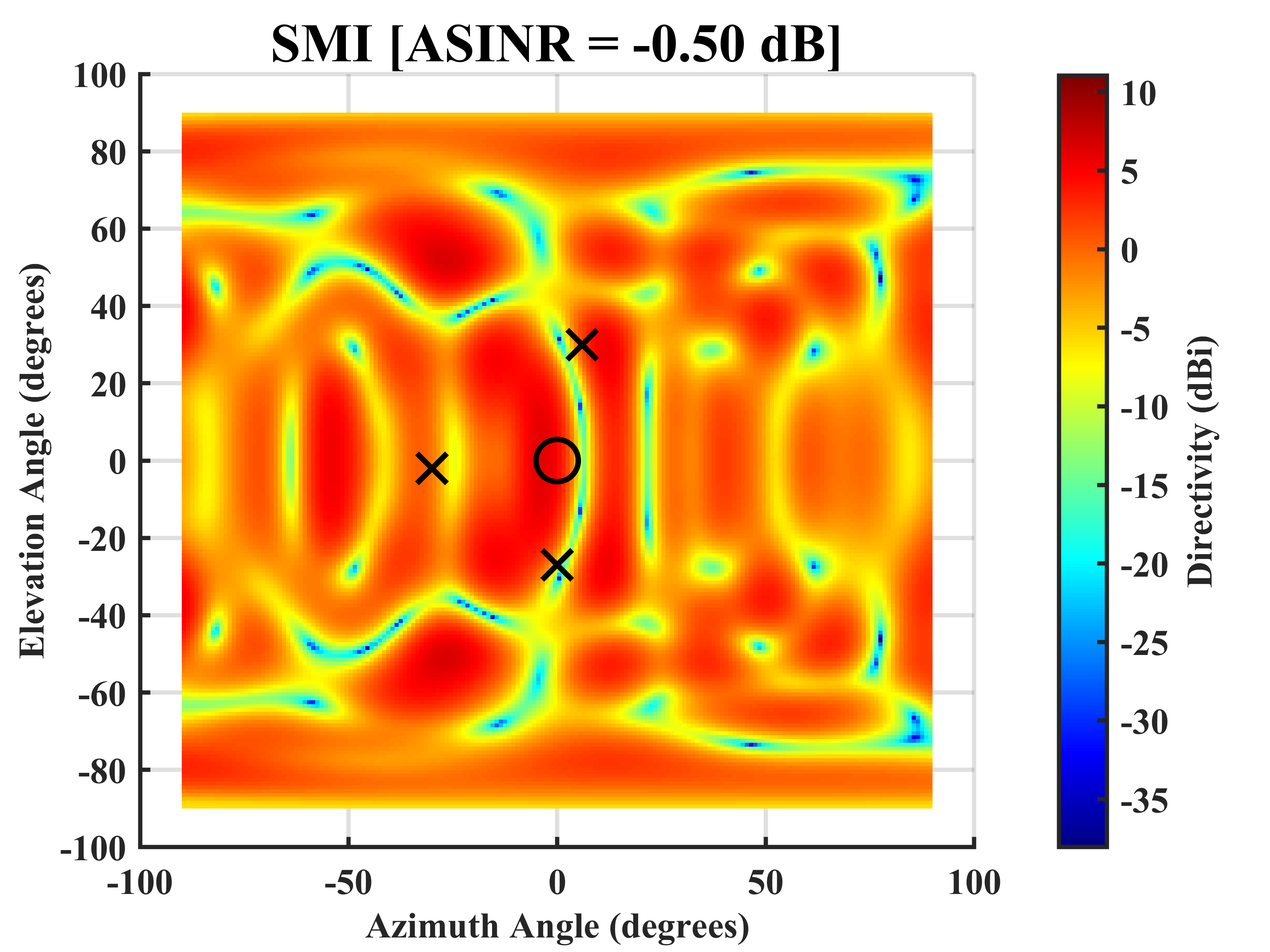}
  \includegraphics[width=2.0in]{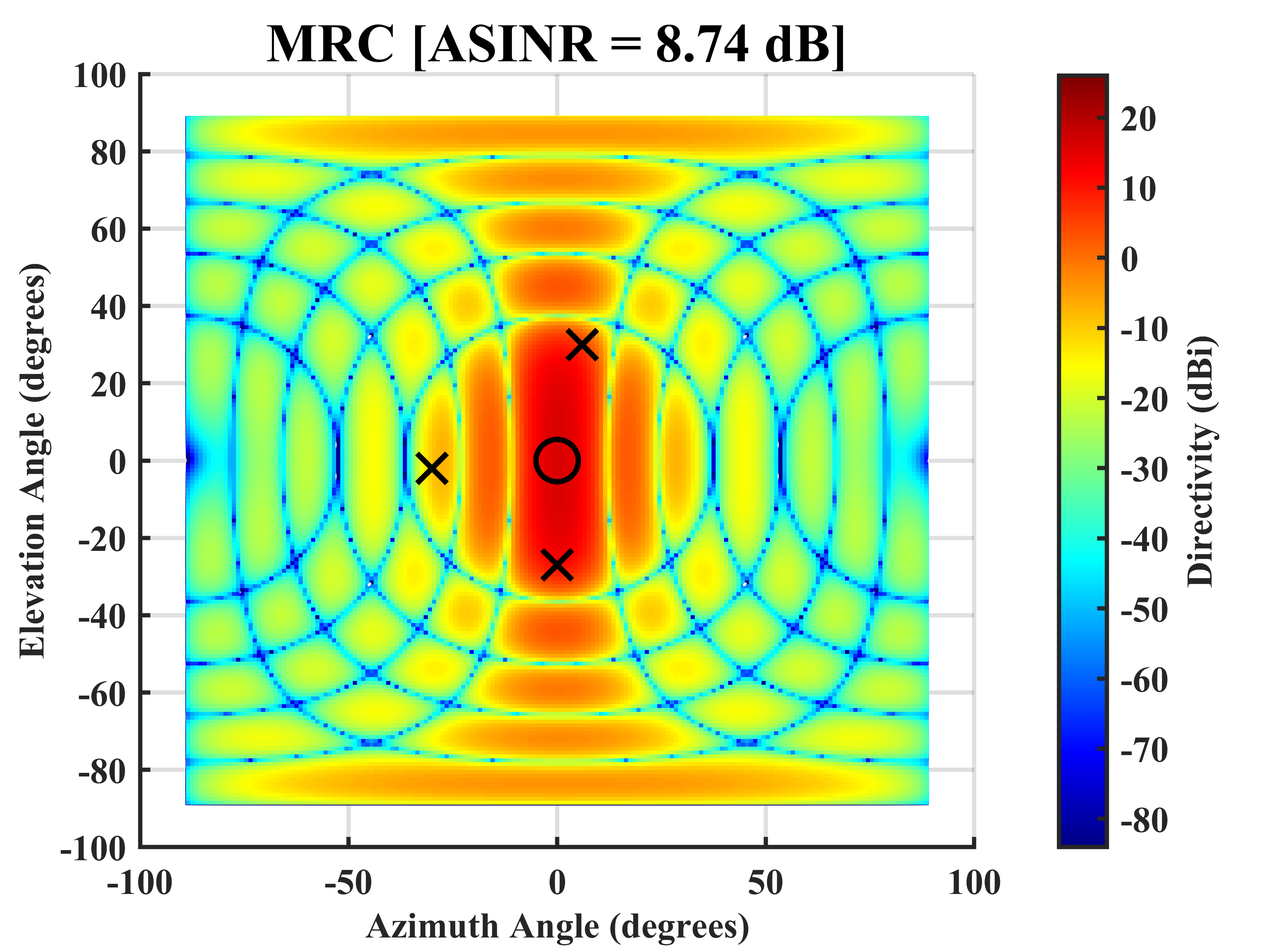}
  \includegraphics[width=2.0in]{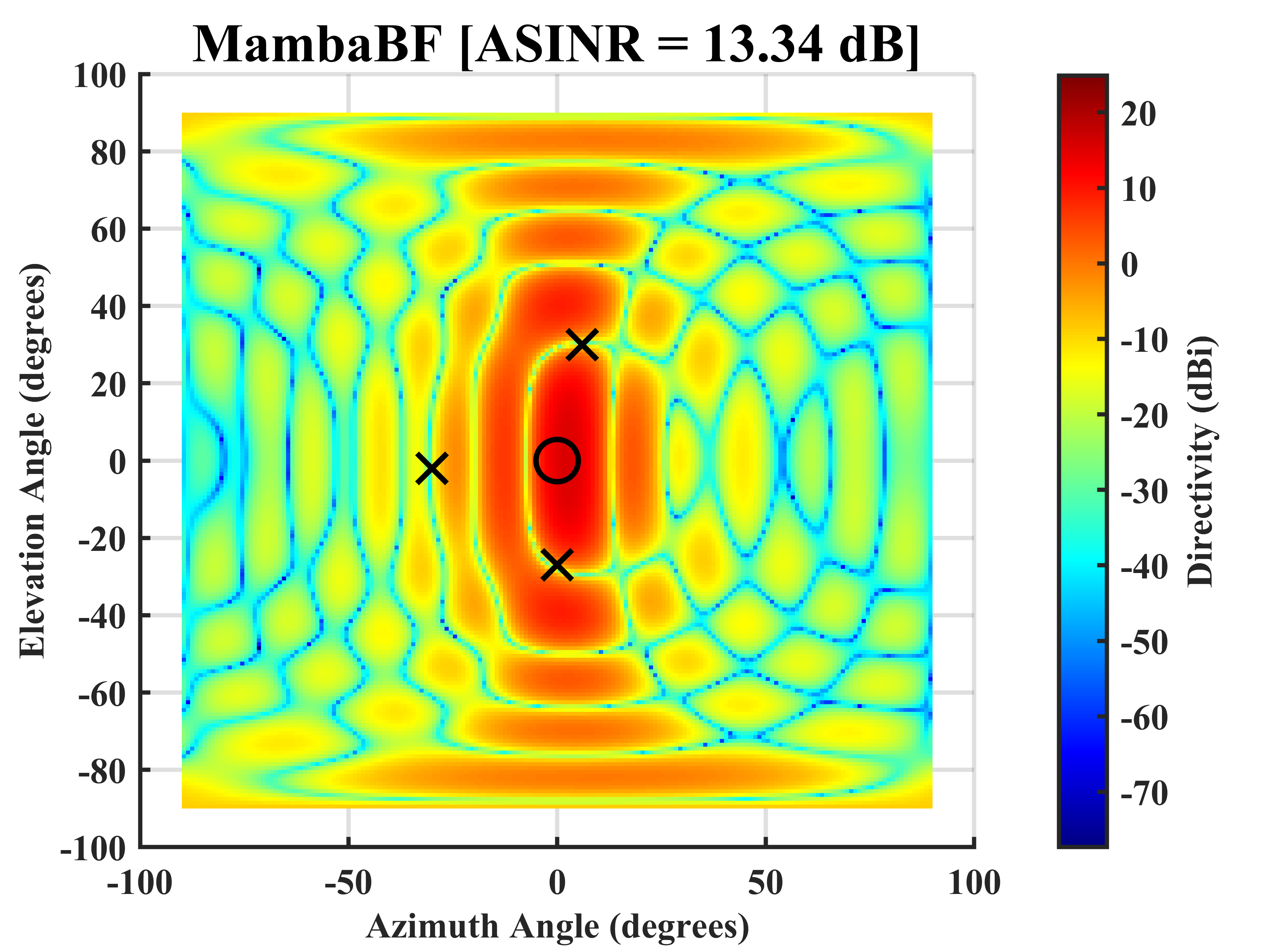}
  \caption{Beam nulling performance (representative case 1: perfect CSI)}
  \label{fig:WPerform1}
\end{figure}

\begin{figure}[ht]
  \centering
  \includegraphics[width=2.0in]{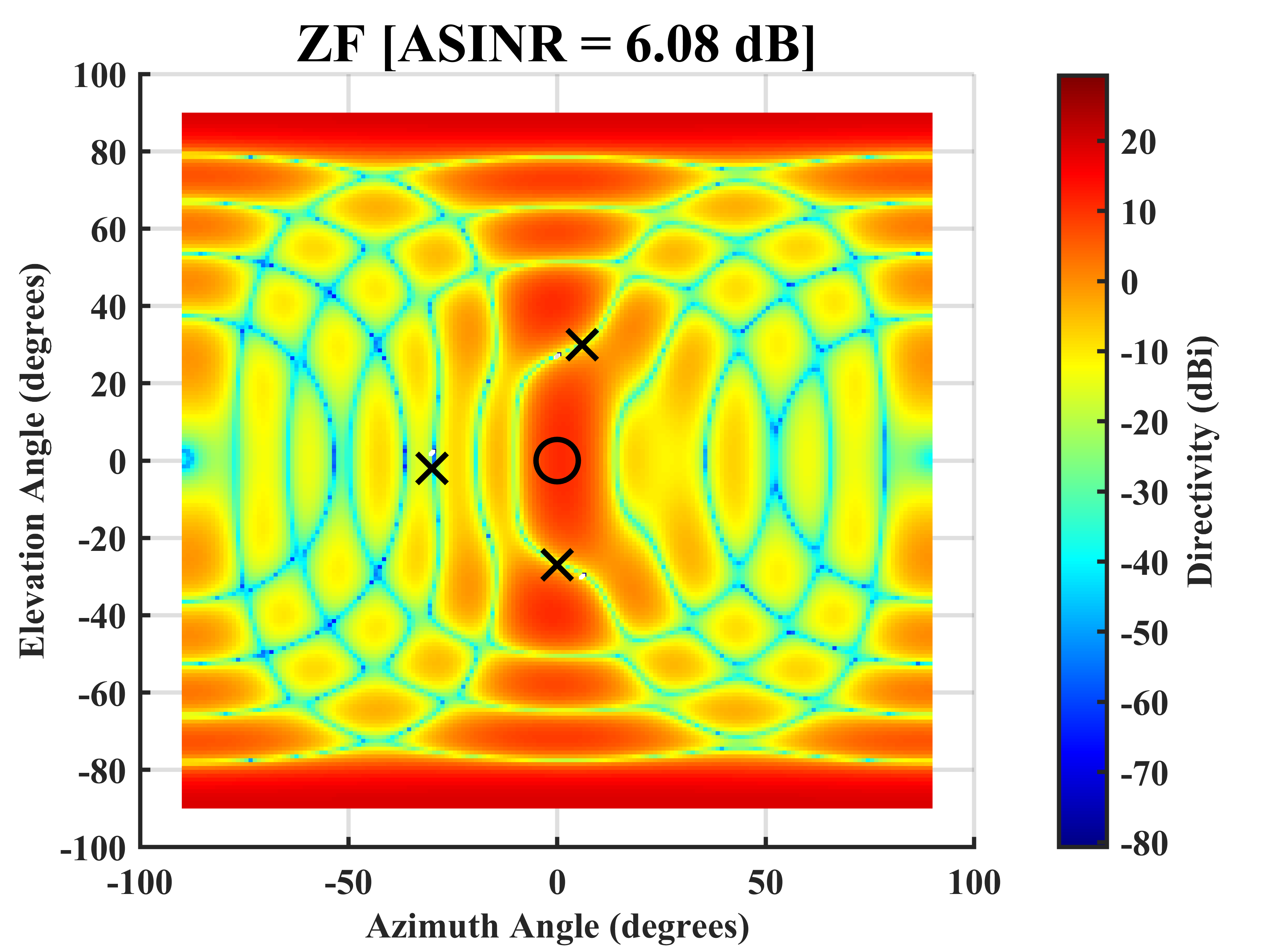}
  \includegraphics[width=2.0in]{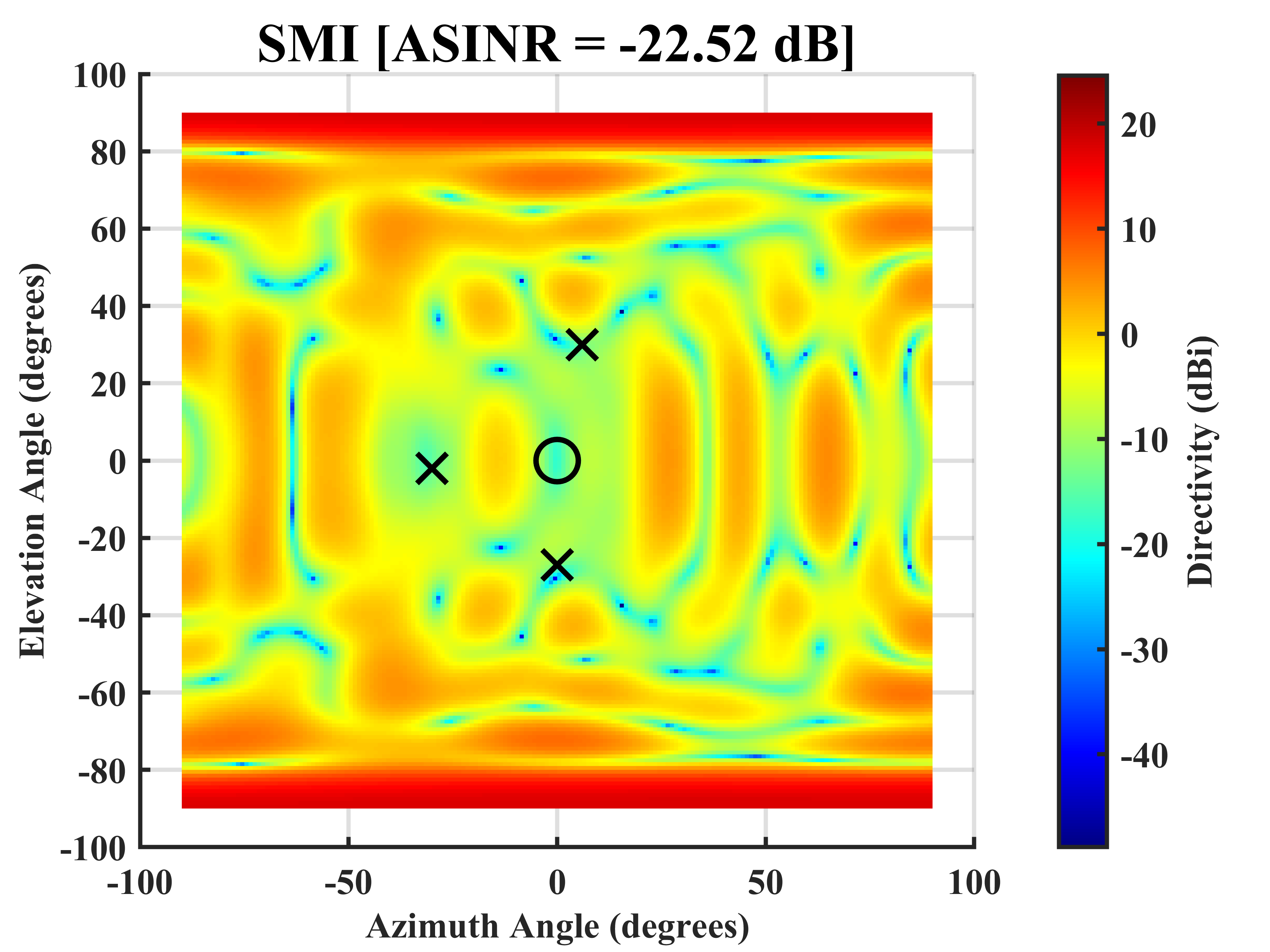}
  \includegraphics[width=2.0in]{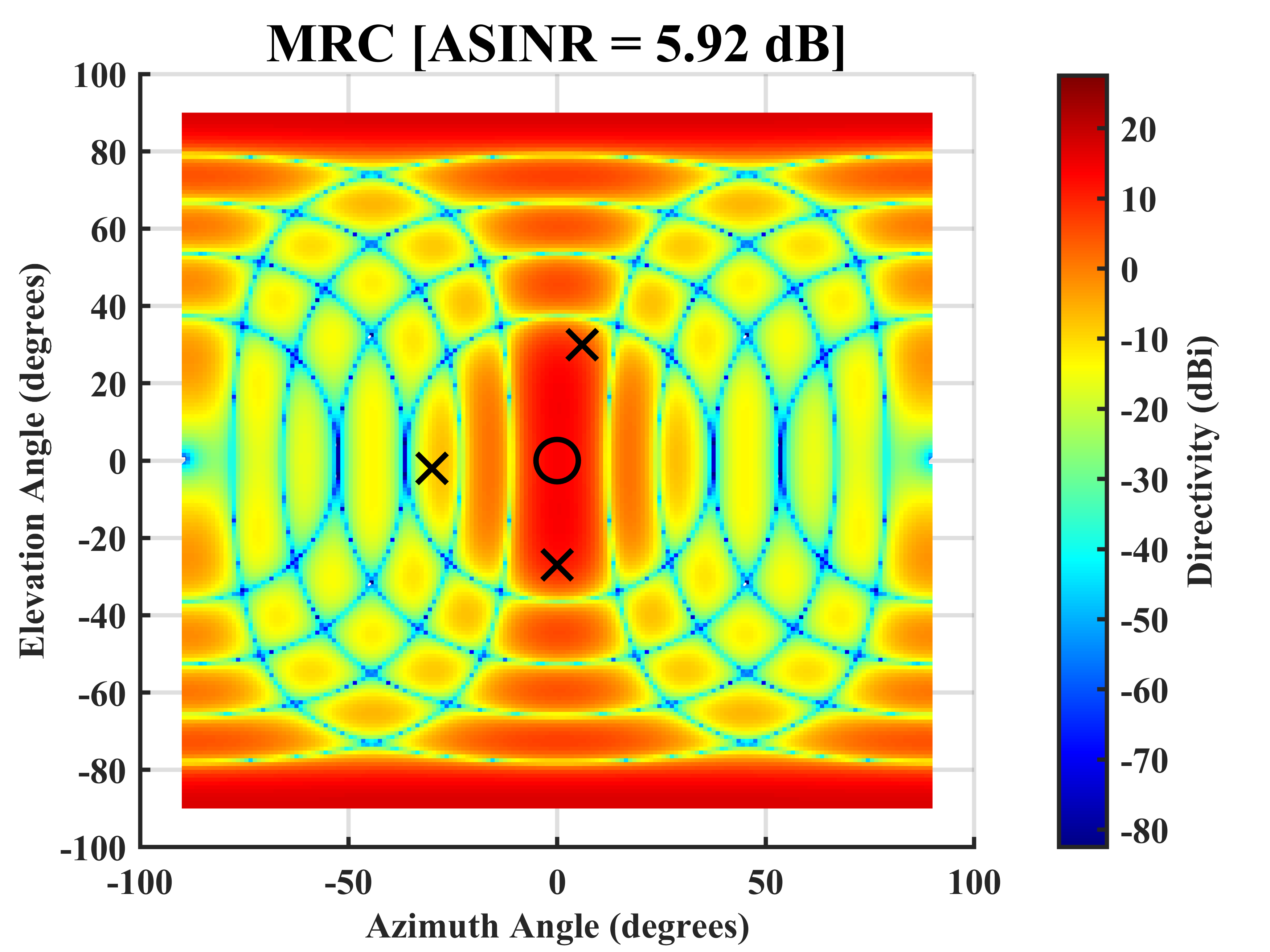}
  \includegraphics[width=2.0in]{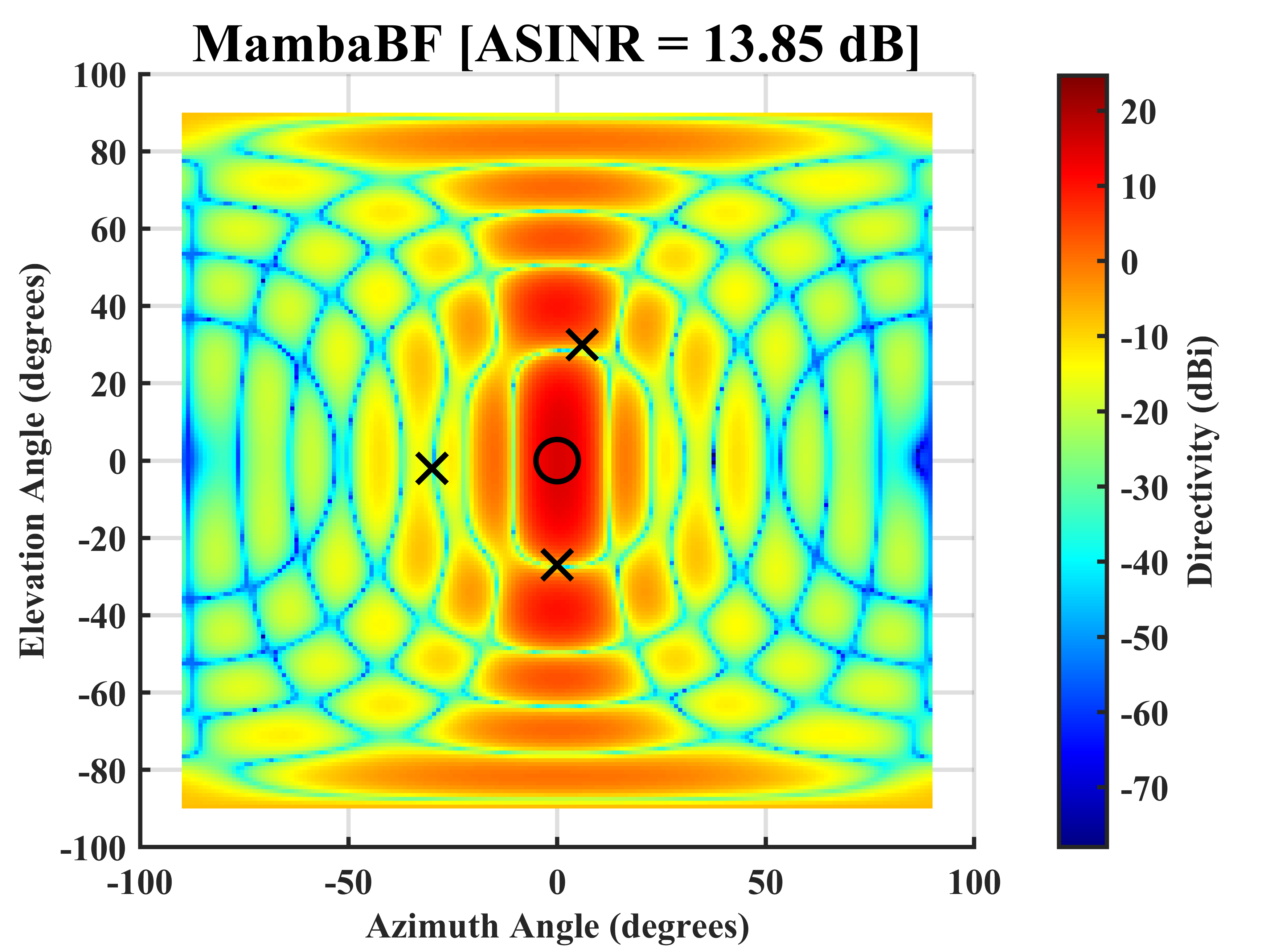}
  \caption{Beam nulling performance (representative case 2: imperfect CSI)}
  \label{fig:WPerform2}
\end{figure}

\subsection{Parameters and Data Generation}
\label{sectionIV:A}

Table~\ref{tab:SmSetup} parameters were used to generate interference scenarios in $\texttt{MATLAB}$, in addition to these parameters, for simplification, the UPA is tilted to capture the $\mathrm{DS}$ signal from DOAs $\left[\phi_d, \theta_d \right] = \left[ 0 , 0 \right] $, while we assume that $\mathrm{ISs}$ transmit severe interference (QoS degradation) could be around side lobe levels (SLLs), so for an array operating in scan angles $\phi, \theta \in  \left[ -90 : 90 \right]$ the critical angles are assumed to be $\phi_{k}, \theta_{k} \in  \left[ -40 : 40 \right]$. The UT directivity in (\ref{eq:ArrayGain}) points at $\mathrm{DS}$ direction with efficiency $\eta^{ut} = 0.99$ and $\mathbf{v}_d$ initial weights. This setup produces different $\mathrm{SINR}$ values, particularly low $\mathrm{SINR}$ at highly correlated desired and interference channels (in-line DOAs angles), and higher $\mathrm{SINR}$ values at less correlated channels or at nulled SLLs locations. The multi-snapshot data is generated at a sampling rate $f_s = 2 \times$ Signal Bandwidth, producing $200$ snapshots per training data sample "$n$", this is for the SMI benchmark where $\mathbf{\hat{R}}$ should be estimated assuming a full rank to ensure a good performance of the matrix inversion. 
\begin{table}[ht]
\caption{Scenarios Parameters}
\label{tab:SmSetup}
\begin{center}
\begin{tabular}{|l|c|}
\hline
\cline{1-2} 
 \textbf{{Parameters}}& \textbf{{Values}} \\
 \hline
 Number of elements ($M$) & $10 \times 10$  elements\\
 \hline
 Number of interfering satellites ($K$)  & 3 satellites \\
 \hline
Satellites altitude ($r_{d}$ , $r_{k}$) &  1000, $\left[500 ~\text{to}~ 600 \right]$ Km \\
 \hline
Satellites EIRP (${P}_{d}$ , ${P}_{k}$) &  45 dBW, $40$ dBW\\
  \hline
Signal Bandwidth ($s_x$,  $s_{i,k}$)  &  50 MHz\\
\hline
Signal Downlink Frequency ($f_{c, d}$ , $f_{c, i}$) &  11.750 GHz\\
\hline
Signal Modcods ($\mathrm{DS}$ \& $\mathrm{ISs}$)  &   QPSK, 8QAM \\
\hline
Number of snapshots ($L$) & $200$ snapshots \\
\hline
 UT equivalent noise temperature ($T^{ut}$) & 230$^{\mathrm{\circ}}$ K \\
 \hline
\end{tabular}
\end{center}
\end{table}

To account for imperfect CSI, we introduce a 15\% misalignment error (i.e., $\sigma_e^2 = 0.15)$ to the desired channel vector $\mathbf{h}_d$, while keeping the interference channel matrix $\mathbf{H}_i$ unaltered.

\subsection{Model Training}
The proposed model processes input tensors of shape $(\text{\text{Batch}}, 2M, L)$, where $\text{\text{Batch}} \leq [N_{\mathrm{tr}}, N_{\mathrm{ts}}]$. We set $N_{\mathrm{tr}} = 4000$ for training and $N_{\mathrm{ts}} = 1000$ for testing. These samples are iteratively and randomly generated using the parameters in Section~\ref{sectionIV:A}.  Throughout the experiments, the \text{Batch} size is fixed to 16, and model weights are updated over 30 epochs. The model reduces the dimensionality of the input to $(\text{\text{Batch}}, M_z, L)$, where $M_z = 100$.

\subsection{Model Performance }

We evaluate the performance of the proposed \textit{MambaBF} beamformer using off-training (test) samples. Earlier, we assumed UPA has knowledge of the $\mathrm{DS}$'s DOA, thus the initial beamformer $\mathbf{w}_{\text{in}}$ is set to $\mathbf{v}_{d}(\varphi_d)$, thus the measured SINR using the initial beamformer will be indicated by $\mathrm{SINR}_{\text{in}}$ in this paper. Our goal is to compare various beamformers' performances by replacing $\mathbf{w}_{\text{in}}$ with a new updated beamformer $\mathbf{w}_{\text{out}}$ and measuring the resulting $\mathrm{SINR}_{\text{out}}(\mathbf{w}_{\text{out}})$ where $\mathbf{w}_{\text{out}} \in [\mathbf{w}_{\mathrm{ZF}}, \mathbf{w}_{\mathrm{MRC}}$ $, \mathbf{w}_{\mathrm{SMI}},\mathbf{w}_{\mathrm{MBF}} ]$.



 

Figs.~\ref{fig:MS_PImCSI} plot $\mathrm{SINR}_{\mathrm{in}}$ versus $\mathrm{SINR}_{\mathrm{out}}$ for both perfect CSI and imperfect CSI conditions. a) Perfect CSI (Left Figure): SMI exhibits poor performance when the number of snapshots (here 200) is limited \footnote{The performance of SMI can improve and get closer to optimal MVDR if the number of snapshots is significantly increased (e.g., 100,000 snapshots), which is substantially higher than the 200 snapshots used for MambaBF.}. ZF performs exceptionally well under high-quality CSI by exploiting full knowledge of the desired and interference channels; it fully cancels interference but struggles in negative $\mathrm{SINR}_{\mathrm{in}}$ due to the high correlation between desired and interference channels (i.e.,~$\mathbf{v}_d \approx \mathbf{v}_{k}$). MRC performs the worst among the classical approaches since it maximizes only the desired signal power without explicitly nulling interference. MambaBF mostly outperforms MRC and sometimes outperforms ZF at low $\mathrm{SINR}_{\mathrm{in}}$, converging at higher $\mathrm{SINR}_{\mathrm{in}}$ with minor fluctuations. b) Imperfect CSI (Right Figure): With channel estimation errors, SMI, ZF, and MRC see noticeable performance drops, illustrating their sensitivity to imperfect CSI.  The MambaBF model, in contrast, outperforms these methods across both low and high $\mathrm{SINR}_{\mathrm{in}}$, showing robustness under realistic imperfect CSI conditions.

Figs.~\ref{fig:WPerform1} and Figs.~\ref{fig:WPerform2} illustrate a representative beam-nulling case under perfect and imperfect CSI conditions, respectively.  Each sub-figure shows the array directivity (in dB) as a function of UT's azimuth and elevation angles. The desired satellite's DOA is marked with a black circle ($\circ$), while the interference satellite's DOAs are indicated by black crosses ($\times$). The average output $\mathrm{SINR}$ ($\mathrm{ASINR}$) for each method is noted in square brackets. Due to the limited space, we cannot cover other cases. These figures show each approach performs at nulling interference locations, demonstrating the strengths and weaknesses of each beamforming approach. With accurate channel knowledge (Fig.~\ref{fig:WPerform1}), ZF effectively cancels the interference sources, achieving high $\mathrm{ASINR}$ at the cost of slightly reduced main-lobe gain in $\mathrm{DS}$ direction. SMI, by contrast, struggles to null both interference signals, leading to negative $\mathrm{ASINR}$ when the number of snapshots is insufficient. MRC maximizes the gain at $\mathrm{DS}$ direction, but does not suppress interference, resulting in a moderate $\mathrm{ASINR}$. 

Notably, MambaBF demonstrates a balanced capability, its beam pattern is directed toward the $\mathrm{DS}$ while also trying to null the interferers, yielding an $\mathrm{ASINR}$ comparable to ZF but without requiring explicit CSI knowledge. Under channel estimation errors, ZF and MRC exhibit performance degradation relative to the perfect-CSI scenario.  SMI is especially sensitive to imperfect covariance estimates, showing a pronounced drop in $\mathrm{ASINR}$. In contrast, MambaBF maintains robust interference nulling and $\mathrm{DS}$ gain enhancement, achieving the highest $\mathrm{ASINR}$ among the methods shown. Because MambaBF does not rely on explicit CSI, it adapts more effectively to channel misalignment errors, confirming its suitability for real-world applications where perfect CSI is rarely guaranteed.

\section{CONCLUSION AND FUTURE WORK}
We have presented \textit{MambaBF}, an self-supervised DL beamforming method that maximizes the SINR and suppresses interference using only the available array snapshots. Numerical results confirm that MambaBF is the most flexible approach, especially under imperfect CSI, where classical methods deteriorate.  This highlights MambaBF’s potential for real-world deployments with inevitable channel imperfections. Potential future directions include evaluating MambaBF under more complex scenarios with moving users, as well as exploring larger-scale datasets for improving the beamforming capability to ZF (optimal) level performance.



\bibliographystyle{IEEEtran}
\bibliography{references}

\end{document}